\documentclass{PoS}
\usepackage[percent]{overpic}

\title{Probing the QCD phase diagram with measurements of strange
  hadron elliptic flow in STAR}

\ShortTitle{Strange hadron elliptic flow in STAR}

\author{\speaker{Md. Nasim (for the STAR collaboration)}\\
  Department of Physics $\&$ Astronomy\\
        University of California, Los Angeles\\
CA-90095, USA\\
        E-mail: \email{nasim@rcf.rhic.bnl.gov}}

\abstract{We present the measurements of strange hadron elliptic flow  at mid-rapidity in Au + Au
  collisions at $\sqrt{s_{NN}}$ = 7.7 - 200 GeV 
  using the STAR detector in the years 2010 and 2011. The transverse
  momentum and collision centrality dependence of elliptic flow  is presented. At the intermediate
  transverse momentum $\Omega$ baryon and $\phi$-meson show 
  baryon-meson separation effect similar to proton and pion for minimum-bias
  Au+Au collision at  $\sqrt{s_{NN}}$ =200 GeV. This indicates
  formation of collective flow at the early partonic phase. 
 The separation between baryons and mesons at intermediate
  transverse momentum decreases with decrease in beam energy and
  almost disappears at $\sqrt{s_{NN}}$ $\leq$ 11.5 GeV, indicating
  hadronic interaction being dominant at the lower beam energy. We
  observe  difference in elliptic flow between particle and
  anti-particle and this increases with decrease in beam
  energy. Differences are larger for baryons than mesons.
  Relative difference between particle and anti-particle
  elliptic flow is larger in central collisions than in peripheral ones.
}

\FullConference{7th International Conference on Physics and Astrophysics of Quark Gluon Plasma,\\
		1-5 February , 2015 \\
		Kolkata, India }

\begin{document}

\section{Introduction}
According to quantum chromodynamics (QCD)  at
very high temperature and/or at high density, a de-confined
phase of quarks and gluons is expected to be present,
while at low  temperature and low density the quarks and gluons
are known to be confined inside hadrons. The aim of the STAR
experiment at Relativistic Heavy Ion Collider (RHIC) is to study the
QCD matter by colliding  nuclei at ultra-relativistic speeds~\cite{white}. Using the information
carried by freely streaming final-state particles as probes, we try to understand the
properties of the medium created in these collisions.
Strange particle elliptic flow ($v_{2}$)  is one of the observables that
is expected to deliver detailed information on the reaction
dynamics of relativistic nucleus-nucleus collisions~\cite{star_v2_auau,star_v2_cucu,alice_v2}.\\
In high energy heavy-ion collisions, particles are produced with an
azimuthally anisotropic momentum distribution.  Elliptic flow  is a measure of  azimuthal angle ($\varphi$) anisotropy of
the produced particles with respect to the reaction plane angle ($\psi$).
The elliptic flow is believed to arise due to the pressure gradients
developed in the overlap region of the two nuclei
colliding at nonzero impact parameter. The $v_{2}$ is an early time
phenomenon and expected to be sensitive to the equation of state
of the system formed in the collision~\cite{hydro}. Thus $v_{2}$ can be used as a probe for
early system although its magnitude may change due to later-stage hadronic
interactions. The interaction cross-sections
of the multi-strange hadrons ($\Xi$, $\Omega$) and $\phi$-meson  with non-strange hadrons are expected to have a small value  and therefore its
production should be less affected by the later stage hadronic interactions in the evolution of the
system formed in heavy-ion collisions~\cite{smallx,BN,NBN}. Moreover they seem to freeze-out early
than non-strange hadrons~\cite{early}. Therefore, multi-strange hadrons can be considered as a clean probe to study the QCD matter.

\section{Data sets and methods}
\begin{figure}[!ht]
\begin{center}
\includegraphics[angle=270,scale=0.55]{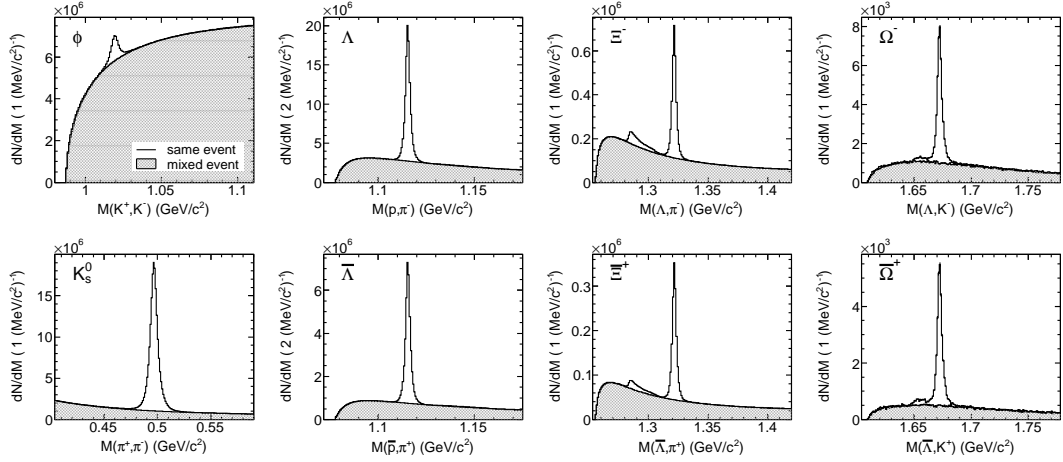}
\caption{(Color online)  Examples of the invariant mass distributions
  at  $\sqrt{s_{NN}}$ = 62.4 GeV for $\phi$, $K^{0}_{S}$,
$\Lambda$($\overline{\Lambda}$), $\Xi^{-}(\overline{\Xi}^{+})$ and
$\Omega^{-}(\overline{\Omega}^{+})$. The combinatorial
background is described by the mixed-event technique. }
\label{msh}
\end{center}
\end{figure}

The results presented here are based on data collected at
$\sqrt{s_{NN}}$= 7.7, 11.5, 19.6, 27, 39,  62.4 and 200 GeV in Au+Au collisions by the
STAR detector for a minimum bias trigger in the years of
2010 and 2011.
The Time
Projection Chamber (TPC)
and Time of Flight (TOF) detectors 
with full $2\pi$ coverage are used for particle identification in the
central pseudo-rapidity ($\eta$) region ($|\eta|<$ 1.0). 
We reconstruct short-lived  $K^{0}_{S}$,
$\Lambda$($\overline{\Lambda}$), $\Xi^{-}(\overline{\Xi}^{+})$, $\Omega^{-}(\overline{\Omega}^{+})$ 
and $\phi$ through the following decay channels :
$K^{0}_{S}$ $\rightarrow$ $\pi^{+}$ + $\pi^{-}$, $\Lambda$
$\rightarrow$ $p$ + $\pi^{-}$ ($\overline{\Lambda}$ $\rightarrow$ $\bar{p}$ + $\pi^{+}$ ), $\Xi^{-}$ $\rightarrow$ $\Lambda$ +
$\pi^{-}$ ($\overline{\Xi}^{+}$ $\rightarrow$ $\overline{\Lambda}$ + $\pi^{+}$), 
$\Omega^{-}$
$\rightarrow$ $\Lambda$ + $\it{K}^{-}$
($\overline{\Omega}^{+}$$\rightarrow$ $\overline{\Lambda}$ +
$\it{K}^{+}$) and $\phi$ $\rightarrow$ $\it{K}^{+}$ +
$\it{K}^{-}$. 
Mixed event technique has been used for combinatorial
background estimation~\cite{phi_plb_star} as shown in Fig~\ref{msh}. The $\eta$-sub event plane
method~\cite{method} using TPC tracks
has been applied to measure the elliptic flow.
In this method, one defines the event flow vector for each
particle based on particles measured in the opposite hemisphere in
pseudo-rapidity ($\eta$).  
\begin{equation}
v_{2}(\eta_{\frac{+}{}}) =
\frac{<\rm{cos}[2(\phi_{\eta_{\frac{+}{}}} -
  \psi_{2,\eta_{\frac{}{+}}})]>}{\sqrt{<\rm{cos}[2(\psi_{2,\eta_{+}}
    - \psi_{2,\eta_{-}})]>}}.
\end{equation}
Here $\psi_{2,\eta_{+}} ( \psi_{2,\eta_{-}} )$ is the second harmonic
event plane angle defined for particles with positive(negative) pseudo-rapidity.
An $\eta$ gap of $\Delta\eta$= 0.1 between positive and negative
pseudo-rapidity sub-events has been introduced to suppress non-flow
effects.

\section{Results}

\subsection{Elliptic Flow at Top RHIC Energy}
The number of constituent quarks (NCQ) scaling in $v_{2}$ for different identified hadrons has been
considered as a good probe for studying the strongly interacting
partonic matter. 
The observed NCQ scaling of identified hadrons in experimental data
~\cite{starphiflow} indicates the importance of parton recombination in forming
hadrons in the intermediate $p_{T}$ range (2.0 GeV/c $<$ $p_{T}$ $<$ 4.0 GeV/c) ~\cite{voloshin,Fries,LXHan}. Such scaling
may indicate that collective elliptic flow is  developed during the
partonic phase. The large statistics data
 sets collected by STAR detectors allow us to measure  elliptic flow of multi-strange
hadrons, specifically that of the $\Omega$ baryon which is made of pure strange ($\it s$) or
anti-strange ($\bar{\it s}$) constituent quarks and of the $\phi$ meson, consisting of one $\it s$  and one
$\bar{\it s}$  constituent quark.   
\begin{figure}[!ht]
\begin{center}
\includegraphics[scale=0.6]{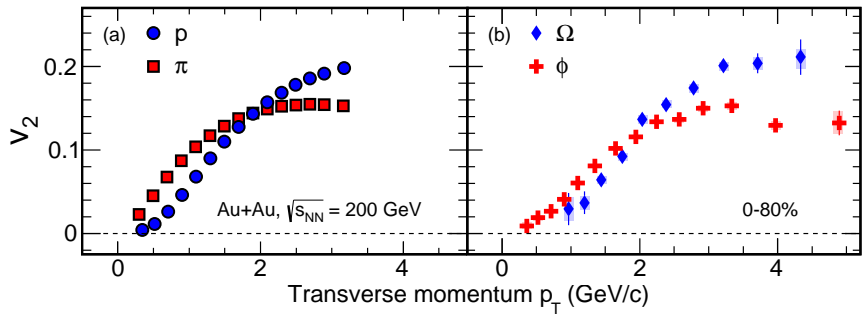}
\caption{(Color online)  The $v_{2}$ as function of $p_{T}$ for
    $\pi$, $p$ (panel a) and $\phi$, $\Omega$ (panel b) from minimum bias Au+Au 
    collisions at $\sqrt{s_{NN}}$ = 200 GeV for 0-80$\%$ centrality~\cite{200GeV_run11_star}. The systematic
    uncertainties are shown by the shaded boxes while vertical lines represent the statistical uncertainties.}
\label{phi_omega}
\end{center}
\end{figure} 
Fig.~\ref{phi_omega} shows the $v_{2}$ as a function of $p_{T}$ for $\pi$, $p$, $\phi$ and $\Omega$ for 0-80$\%$ centrality in Au + Au collisions at
$\sqrt{s_{NN}}$ = 200 GeV~\cite{200GeV_run11_star}. 
Fig.~\ref{phi_omega} (a) shows a comparison between
$v_{2}$ of $\pi$ and $p$, consisting of up ($\it u$) and down ($\it d$) light quarks, and
Fig.~\ref{phi_omega} (b) shows a comparison of $v_{2}$ of $\phi$ and $\Omega$
containing  heavier $\it s$ quarks. 
The $v_{2}$ of  $\phi$ and $\Omega$ are mass
ordered  at low $p_{T}$ and a baryon-meson separation  is observed at intermediate $p_{T}$.
It is clear from Fig.~\ref{phi_omega} that the $v_{2}(p_{T}) $ of hadrons consisting only of strange quarks ($\phi$ and $\Omega$)  is similar to that of $\pi$ and $p$. However, unlike $\pi$ and $p$, the $\phi$ and
$\Omega$ do not participate strongly in the hadronic interactions, which suggests that the major part of collectivity is developed during the
partonic phase in Au + Au collisions at $\sqrt{s_{NN}}$ = 200 GeV.\\

\begin{figure}[!ht]
\centering
\centerline{\includegraphics[scale=0.7]{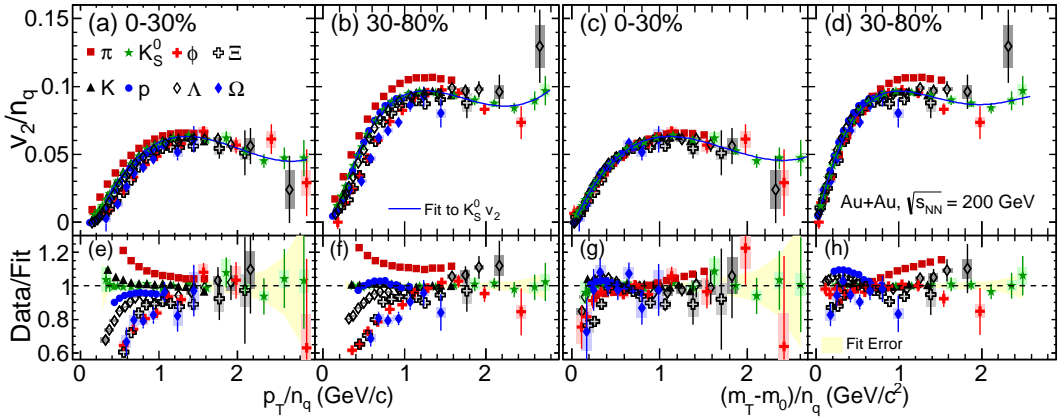}}
  \caption{(Color online) The $v_{2}$ scaled by number of constituent
    quarks ($n_{q}$) as a function of  $p_{T}/n_{q}$ and  $(m_{T}
    -m_{0})/n_{q}$ for identified hadrons from Au + Au collisions at
    $\sqrt{s_{NN}}$ = 200 GeV~\cite{200GeV_run11_star}. Ratios with respect to a fit to the
    $K^{0}_{S}$ $v_{2}$ are shown in the corresponding lower
    panels. Vertical lines are  statistical uncertainties  and shaded boxes are
    systematic uncertainties.}
\label{phi_ncq}       
\end{figure}

Fig.~\ref{phi_ncq} shows the $v_{2}$ scaled by number of
constituent quarks ($n_{q}$) as a function of $p_{T}/n_{q}$ and $(m_{T}-m_{0})/n_{q}$ for
identified hadrons from Au + Au collisions at  $\sqrt{s_{NN}}$ = 200 GeV 
for 0-30$\%$ and 30-80$\%$ centrality~\cite{200GeV_run11_star}, where $m_{T}$ and $m_{0}$ are
the transverse mass and rest mass of  hadron, respectively. To quantify the deviation from
NCQ scaling, we fit the $K^{0}_{S}$ $v_{2}$ with a third-order
polynomial function.  We then take the ratio of $v_{2}$ for the other
measured hadrons to the $K^{0}_{S}$ fit.  The ratios are shown in the
lower panels of Fig.~\ref{phi_ncq}. For both 0-30$\%$ and 30-80$\%$ centralities, the scaling holds approximately
 within 10$\%$, excluding pions. The deviation of pions could be due
 the effect of resonance decay and non-flow correlations~\cite{dongx}. 

\subsection{Elliptic Flow  at Beam Energy Scan}
\subsubsection{Elliptic flow as a function of transverse mass}
\begin{figure}[!ht]
\begin{center}
\centerline{\includegraphics[scale=0.7]{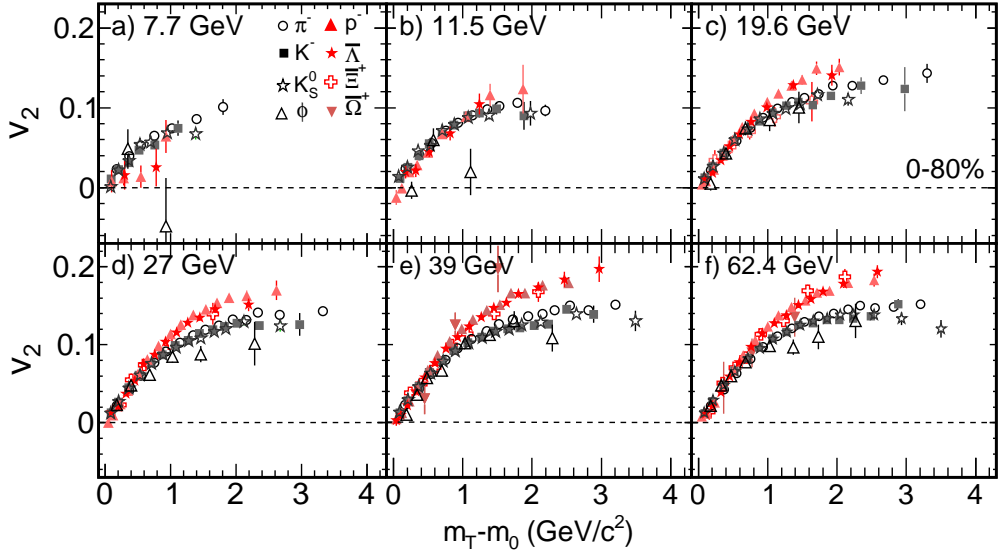}}
\caption{(Color online) The elliptic flow ($v_{2}$) as a
  function of $(m_{T}-m_{0})$ for selected particles in the Au+Au collision
  at various beam energies for 0-80$\%$ centrality~\cite{prc,prl}. Error bars are only
  statistical uncertainties.}
\label{bm_bes}
\end{center}
\end{figure}

Fig.~\ref{bm_bes} shows $v_{2}$ as function  of  $(m_{T} - m_{0})$
at $\sqrt{s_{NN}}$ = 7.7-62.4 GeV~\cite{prc,prl}. 
There is a clear splitting between baryons and mesons for larger
$(m_{T} - m_{0})$ values at $\sqrt{s_{NN}}$ = 62.4 GeV. As we go down
in energy, the splitting becomes narrower and at 11.5 GeV, the difference between the baryons and mesons is no longer
observed. Also, we observed that $v_{2}$ of $\phi$ mesons falls off the
trend from the other hadrons at  $\sqrt{s_{NN}}$ $\leq$ 11.5 GeV. This could be related to the lower hadronic
cross sections of particles containing multiple strange quarks~\cite{smallx,BN,NBN}.
These observations may indicate that hadronic interactions
become more important than partonic ones for the systems
formed at collision energies $\sqrt{s_{NN}}$ $\leq$ 11.5 GeV.

\subsubsection{Energy and centrality dependence of the particle
and antiparticle $v_{2}$ difference}
\begin{figure}[!ht]
\begin{center}
\begin{overpic}[scale=.7]{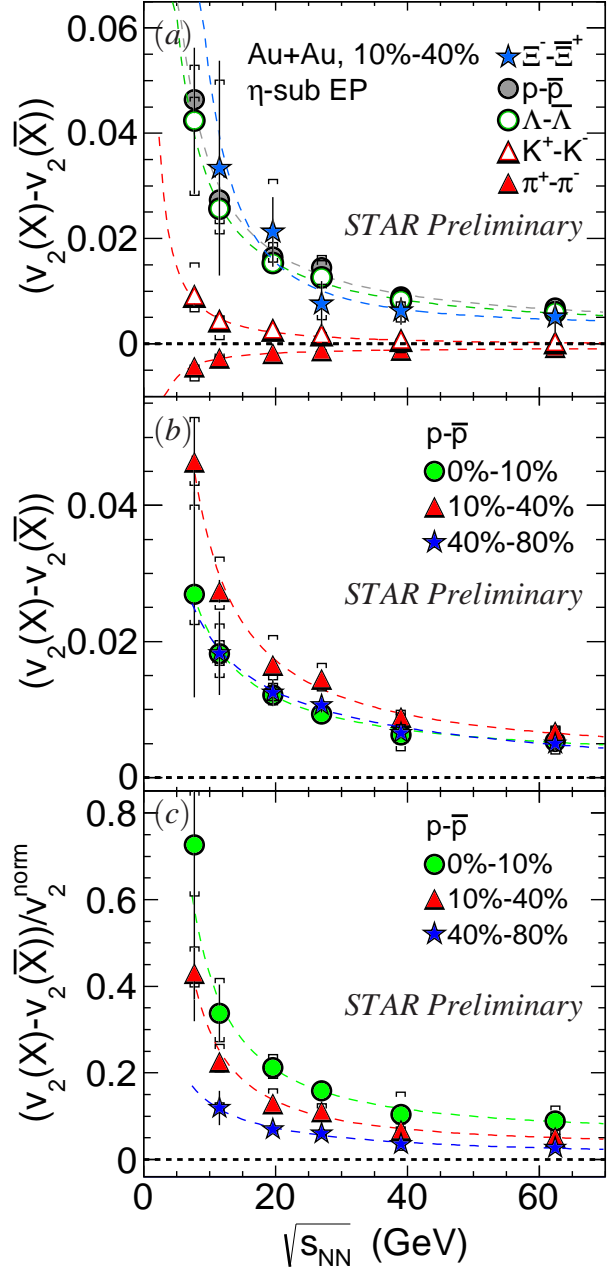}
\put (25,50) {\large $STAR$  $Preliminary$}
\put (25,20) {\large $STAR$  $Preliminary$}
\put (25,77) {\large $STAR$  $Preliminary$}
\put (11,91) {\large $(a)$}
\put (11,62) {\large $(b)$}
\put (11,34) {\large $(c)$}
\end{overpic}
\caption{(Color online) The difference in  $v_{2}$ values between
a particle($X$)  and its corresponding antiparticle ($\overline{X}$) as a function of
$\sqrt{s_{NN}}$. The dashed lines in plot are fits using function
$f_{\Delta v_{2}}(\sqrt{s_{NN}})=a \times s_{NN}^{-b/2}$.  Here
$v_{2}^{\rm{norm}}$ is equal to proton $v_{2}$ at $p_{T}$=1.5
GeV/$c$.}
\label{v2diff_bes}
\end{center}
\end{figure}

The energy dependence of the $v_{2}$ difference
between particles and antiparticles for different centralities are shown in Fig.~\ref{v2diff_bes}. Top panel shows the
differences for $\pi$, $K$, $p$, $\Lambda$ and $\Xi$ at 10-40$\%$
collisions centrality. We can see that the difference for all baryons
are same at 10-40$\%$ centrality for all $\sqrt{s_{NN}}$, which is
consistent with the observation at   0-80$\%$~\cite{prc,prl}. Middle
and lower panel of Fig.~\ref{v2diff_bes} show the differences between
protons and anti-protons for 0-10$\%$, 10-40$\%$
and 40-80$\%$ centralities. The y-axis of the lower panel is scaled by the
proton $v_{2}$ at $p_{T}$=1.5 GeV/$c$ (labeled as
$v_{2}^{\rm{norm}}$) to show the relative difference. We can see, the
relative difference in $v_{2}$ between protons and anti-protons
increases from peripheral (40-80$\%$) to central (0-10$\%$) collisions.
This observation support the model prediction~\cite{hybrid_v2diff}, which includes baryon
stopping as a mechanism to explain the data. 

\section{Summary}
Energy and centrality dependence of strange hadron $v_{2}$ at mid-rapidity are presented. 
The $p_{T}$ dependence of $\phi$ and $\Omega$ 
 $v_{2}$ is similar to $\pi$  and $p$ $v_{2}$ at top RHIC energy, which indicates that the major
parts of collectivity were developed at the initial partonic phase at
$\sqrt{s_{NN}}$ = 200 GeV. To investigate partonic
collectivity for different system size, NCQ scaling has been presented
for two different collision centralities, 0-30$\%$ and  30-80$\%$. It
is observed that the NCQ scaling holds within the statistical
uncertainty for both  0-30$\%$ and  30-80$\%$ centralities. 
Splitting between baryon and meson at intermediate $p_{T}$, which formed the basis of NCQ
scaling observation at top RHIC energy, was not observed  at the lower
energies indicating formation
 of a matter mostly governed by hadronic interaction. 
We observed beam-energy dependent difference in $v_{2}$ between particle and corresponding
anti-particle. Differences are larger for baryons than mesons. The difference increases with decreasing beam
energy. Relative difference between particle and anti-particle $v_{2}$
is larger in central collisions than in peripheral ones.

\section{Acknowledgements}
Financial support from DOE project is gratefully acknowledged. 

\end{document}